\newif\ifPDF
\documentclass[usenatbib]{mn2e}
\ifPDF
  \usepackage[T1]{fontenc}
  \usepackage{times}
  \usepackage[hypertex,colorlinks=true,citecolor=blue]{hyperref}
  \usepackage{aeguill}
  \usepackage[pdftex]{graphicx,color}
  \usepackage{subfigure}
  \usepackage{afterpage}
\else
  \usepackage[T1]{fontenc}
  \usepackage{times}
  \usepackage[dvips]{graphicx,color}
  \usepackage{subfigure}
  \usepackage{afterpage}
\fi
\title[Observations of four glitches in the young pulsar J1833$-$1034 and study of its glitch activity]
{Observations of four glitches in the young pulsar J1833$-$1034 and study of its glitch activity}
\author[Jayanta Roy, Yashwant Gupta \& Wojciech Lewandowski]
{Jayanta Roy$^1$, Yashwant Gupta$^1$, Wojciech Lewandowski$^2$\\\\
$^1$National Centre for Radio Astrophysics, TIFR, Pune University Campus, Post Bag 3,
Pune 411 007, India\\
$^2$Institute of Astronomy, University of Zielona Gora, Lubuska 2, 65-265 Zielona Gora, Poland }

\date{Accepted. Received}
\begin{document}
\label{firstpage}
\maketitle
\pagerange{\pageref{firstpage}--\pageref{lastpage}} \pubyear{2009}
\def\LaTeX{L\kern-.36em\raise.3ex\hbox{a}\kern-.15em
    T\kern-.1667em\lower.7ex\hbox{E}\kern-.125emX}

\begin{abstract}
We present the results from timing observations with the GMRT of the young
pulsar J1833$-$1034, in the galactic supernova remnant G21.5$-$0.9. We detect the presence 
of 4 glitches in this pulsar over a period of 5.5 years, making it one of a set of pulsars 
that show fairly frequent glitches. The glitch amplitudes, characterized
by the fractional change of the rotational frequency, range from $1\times10^{-9}$ to
$7\times10^{-9}$, with no evidence for any appreciable relaxation of the rotational frequency
after the glitches. The fractional changes observed in the frequency derivative are of the order 
of $10^{-5}$. We show conclusively that, in spite of having significant timing noise, the sudden 
irregularities like glitches detected  in this pulsar can not be modeled as smooth timing noise.    
Our timing solution also provides a stable estimate of the second derivative of the pulsar 
spin-down model, and a plausible value for the braking index of 1.857, which, like the value for 
other such young pulsars, is much less than the canonical value of 3.0. 
PSR J1833$-$1034 appears to belong to a class of pulsars exhibiting fairly frequent 
occurrence of low amplitude glitches.  This is further supported by an estimate of the glitch activity 
parameter, $A_{g} ~=~ 1.53\times10^{-15}$ $s^{-2}$, which is found to be significantly lower than the trend of glitch 
activity versus characteristic age (or spin frequency derivative) that a majority of the glitching 
pulsars follow.  We present evidence for a class of such young pulsars, including the Crab, where 
higher internal temperature of the neutron star could be responsible for the nature of the observed 
glitch activity.  
\end{abstract}

\label{firstpage} \pagerange{\pageref{firstpage}--\pageref{lastpage}} %
\pubyear{2011}


\section{Introduction}                \label{sec:intro}    
Beside the basic, smooth spin-down of the neutron star due to the electromagnetic 
torque mechanism, pulsar timing studies also reveal the presence of irregularities
in the rotation of the star, mainly of two kinds : timing noise, which is characterized by continuous, random fluctuations 
in the rotation rate; and glitches, which are sudden increases in the rotation rate. 
 These spin-up events are superposed on the long-term spin-down of the pulsar, and 
manifest themselves as sudden early arrival of the pulses. A recent work by 
\citep{Espinoza11} reports a total of 315 glitches observed in 102 pulsars. The magnitude of the change in rotation frequency, 
$\nu$, during a glitch is typically in the range $10^{-10}$ $<$ $\Delta \nu / \nu $ $<$ $10^{-6}$, 
and the fractional increment in the spin-down rate, $\Delta \dot{\nu} / \dot{\nu}$, is in the range 
$10^{-5}$ to $10^{-2}$. 

The most plausible explanation for the sudden spin-up is the irregular flow of angular momentum from the 
faster rotating superfluid interior to the more slowly rotating solid crust of the neutron star as it slows down 
\citep{Lyne95}. The current unified model for glitches is based on the superfluidity of the neutrons in a neutron star. 
The rotating superfluid in the neutron star carries angular momentum, by forming quantized vortices. The spacing 
between the vortices is negligible compared to the radius of the neutron star. On the macroscopic scale, the flow pattern 
looks like uniform rotation. These quantized vortices in the neutron superfluid in the inner crust can get pinned 
to the lattice of heavy neutron-rich nuclei.
The pinning is possible because the effective width of the vortex core is less than 
or comparable to the lattice spacing of the nuclei. The pinning force is related 
to the energy gain when vortices are pinned to the lattice.  The vortices stay pinned 
in this manner until a stronger force unpins them from the lattice sites. 
These pinned vortices in the crustal nuclei are rotating slower than the surrounding 
superfluid. Due to this differential velocity, magnus forces that act radially outward 
cause sudden unpinning and migration of vortices, which results in the transfer of angular 
momentum from the superfluid to the crust. This gives rise to a sudden speed-up of 
the solid crust, which manifests as a glitch in the timing behaviour of the pulsar. 
\cite{Anderson75} were the first to make this connection between sudden unpinning 
of vortices and pulsar glitches. In the unpinned state, the superfluid
moment of inertia is not coupled to the crust, hence the effective moment of inertia of the crust decreases, 
which in-turn increases the spin-down rate. Between glitches, the vortex lines undergo a slow, thermally 
activated process, called vortex creep. The post-glitch relaxation is a process of recoupling of the 
vortices to another steady  (pinned) state. Once the moment of inertia recovers due to this recoupling 
via repinning, the original extrapolated spin-down rate is restored. Thus, the observed sudden increase 
in the rotation rate, followed by exponential relaxation back to the extrapolated 
pre-glitch rotation rate, provides a useful probe of the neutron star interior.

The pulsar J1833$-$1034 was independently discovered at the GMRT \citep{Gupta05} and at Parkes \citep{Camilo06} 
and is associated with the galactic supernova remnant (SNR) G21.5$-$0.9.  This pulsar has
quite a high spin-down luminosity that is amongst the top ten of all the known pulsars in our Galaxy. 
The flux density measured at radio wavelengths is very low $-$ the estimated mean flux density from the 
observations at 610 MHZ is 0.65 mJy \citep{Gupta05}. With a period of 61.86 ms
and a period derivative of 2.02$\times$10$^{-13}$ s/s, it has a characteristic age, 
$\tau_{c}$  $\approx$ 4.8 kyr \citep{Camilo06}, which makes it a fairly young pulsar. 
Existing studies indicate that younger pulsars are more likely to show glitches.  
About half of all known pulsars with $\tau_{c}$ less than 3$\times$10$^{4}$ have 
exhibited glitches, but this fraction is much lower for the older population \citep{Yuan10}. 
PSR J1833$-$1034 is thus a good candidate for the study of glitches.  In this paper 
we report the detection of multiple glitches from this pulsar using timing observations
carried out at the GMRT at 610 MHz, and present a detailed study of its glitch activity.  
We also provide refined estimates of the timing parameters for this pulsar, including 
an estimate of the braking index. In Section 2 we explain the observations and 
data analysis techniques. Section 3 describes the detected glitches and their modeling 
in detail. In Section 4 we discuss the significance of our results. Summary and future 
scope are presented in section 5.  

\section{Observations and data analysis}  \label{sec:observations} 
The GMRT is a multi-element aperture synthesis telescope consisting of 30 antennae, each of 45 m
diameter, spread over a region of 25 km diameter \citep{Swarup97}.  Though designed to function 
primarily as an aperture synthesis telescope, the GMRT can also be used as an effective single
dish in an array mode by adding the signals from individual dishes, either coherently or incoherently 
\citep{Gupta00}, for studying compact objects like pulsars. The radio signals at the observing frequency 
band, from both polarizations of the 30 dishes, are eventually converted to baseband signals of 
16 MHz or 32 MHz bandwidth, which are then sampled at Nyquist rate.  These digitised signals are delay 
corrected and then Fourier Transformed in a FX correlator to get spectral information.  After fringe 
derotation, these dual polarization spectral voltage samples from all the antennae are added coherently 
in the GMRT Array Combiner (GAC), to produce the phased array outputs for each polarization.  These are 
then converted to intensity, integrated to the desired time constant and recorded 
on disk for off-line processing.  The data are time-stamped using a minute pulse signal, derived from the observatory's GPS receiver, 
which is embedded in the data stream.

The timing observations described here were carried out in the total intensity phased array mode at 
610 MHz. In this mode of operation, the array needs to be phased up before observing the target pulsars. 
This is achieved by recording the correlator data for a point source calibrator, solving for the antenna
based gains and phases from these, and applying the phases as corrections to the output of the Fourier 
Transform stage of the correlator.  The array remains phased for up to a few hours and dephases due to 
slow changes in instrumental and ionospheric phases. When this happens, one needs to rephase the array 
to proceed with further observations. 

%%%%%%%%%%%%%%%%%%%%%%%%%%%%%%%%%%%%%%%%%%%%%%%%%%%%%%%%
\begin{table*}
\begin{center}
\caption{Main parameters for the two pulsars observed, at 610 MHz with a bandwidth 
of 16 MHz, with a typical cadence of 10 days.}
\vspace{0.3cm}
\label{observation_summary}
\begin{tabular}{|l|c|c|c|c|c|c|c|c|c|c|c|c|c|c|c|c}
\hline
PSR     &Period   &Mean flux at  &Integration   &$N_{p}$ &$(S/N)_{exp}$  \\
        &(ms)     &610 MHz(mJy)  &time (min)   &        &               \\\hline
        &         &              &              &        &               \\
B1855+09&5.36     &16.8$^{\dagger}$&5           &55970   &177            \\
J1833-1034&61.86  &0.65$^{\ddagger}$&90         &87293   &32             \\\hline
\end{tabular}
\vspace{0.3cm}\\
\begin{flushleft}
${\dagger}$ extrapolated flux using the catalogued values at 400 and 1400 MHz.\\
${\ddagger}$ \cite{Gupta05}\\ 
$N_{p}$ is number of pulses accumulated in the integration time.\\
$(S/N)_{exp}$ is the expected S/N for the 610 MHz profile at any epoch.
\end{flushleft}
\end{center}
\end{table*}
%%%%%%%%%%%%%%%%%%%%%%%%%%%%%%%%%%%%%%%%%%%%%%%%%%%%%%
Timing observations for PSR J1833$-$1034 were started around mid-2005, shortly after its discovery at 
the GMRT.  In the beginning, after the initial, closely spaced observations that are needed to obtain 
the timing solution for a newly discovered pulsar, the observations were somewhat random and sparse in 
time. Since the occurrences of glitches are unpredictable and their relaxation timescales can be quite 
short, regular monitoring is important for detection and study of the glitches.  From mid-2007, after 
the possible detection of the first glitch from this pulsar, a regular timing program was started, with  
observations roughly about 10 days apart, except for the GMRT maintenance intervals.  Each observing epoch 
has a 90 min long scan on PSR J1833$-$1034, and a shorter scan of 5 min on PSR B1855$+$09, which acts 
as a control pulsar for validating the data quality and reliability of the time-of-arrival (TOA) values 
from the newly established GMRT timing pipeline.  The final data for each scan are total intensity values 
for each of 256 spectral channels (across a 16 MHz bandwidth), recorded with a sampling interval of 0.256 ms.   
The main observing parameters for a typical epoch are summarized  in Table \ref{observation_summary}.

\begin{figure}
\begin{center}
\includegraphics[angle=270,width=0.5\textwidth]{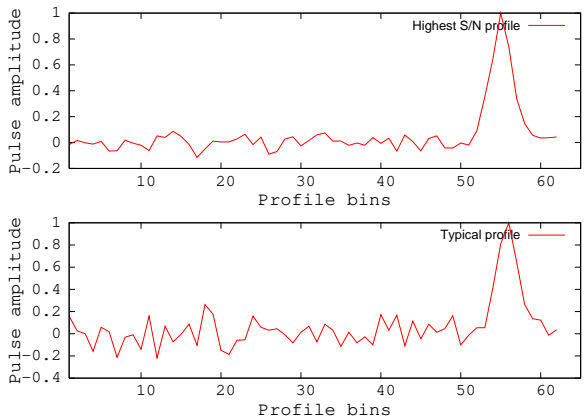}
\caption[Profiles of J1833$-$1034]{Highest signal-to-noise ratio profile (top 
panel) and typical average profile (bottom panel) for PSR J1833$-$1034. The 
highest signal-to-noise ratio is 22, whereas the typical average profile has a 
signal-to-noise ratio of 10.}
\label{J1833-1034_profiles}
\vspace{0.1cm}
%\hrule height 1.5pt
\end{center}
\end{figure}

In the off-line processing, the recorded multi-channel total intensity data are first dedispersed to remove 
the effect of interstellar dispersion on the pulse shape. The dedispersed time series data are then 
synchronously folded using the topocentric pulsar period, obtained from the best existing model parameters 
(barycentric) for the concerned pulsar, after correcting for the observing time and location. The UTC 
corresponding to the middle of the observing session is used as the reference point for that particular 
epoch and it is derived from analysis of the GPS pulse signal embedded in the data. 
The predicted topocentric periods are calculated using  ``polyco'' files produced by the pulsar timing 
program TEMPO \footnote{see http://www.atnf.csiro.au/research/pulsar/tempo}. For the control pulsar, 
the barycentric model parameters were taken from the ATNF pulsar catalog 
\footnote{see http://www.atnf.csiro.au/research/pulsar/psrcat};
for J1833$-$1034, these were obtained from the initial epochs of observations and refined at successive 
epochs, as required.  The topocentric TOAs at each epoch are obtained by cross-correlating the average 
profile at that epoch with the highest signal-to-noise (S/N) profile from all the epochs used as a 
template. These are then converted to solar system barycentric TOAs using the Jet Propulsion Laboratory 
DE200 solar system ephemeris \citep{Standish82} inside TEMPO. To illustrate the quality of the profiles, 
Fig. \ref{J1833-1034_profiles} shows the highest S/N profile (which is used as the template), as well as 
a typical average profile (whose S/N is close the median value from profiles of all epochs).  For the
timing analysis using TEMPO, we have used these topocentric TOAs along with the uncertainties related 
to the S/N of the profiles. The tiny uncertainty of the TOA of the reference epoch is artificially 
increased to make it close to the median value.   
%%%%%%%%%%%%%%%%%%%%%%%%%%%%%%%%%%%%%%%%%%%%%%%%%%%
%%%%%%%%%%%%%%%%%%%%%%%%%%%%%%%%%%%%%%%%%%%%%%%%%%%%

At the solar system barycentre, the time evolution of the rotational phase of a solitary pulsar is 
well-approximated by a polynomial of the form \citep{Manchester77}
\begin{equation}
\phi_{m}(t) = \phi_{0} + \nu (t-t_{0}) + \frac{1}{2} \dot{\nu} (t-t_{0})^{2} + \frac{1}{6} \ddot{\nu} (t-t_{0})^{3}
\end{equation}
where $\phi_{0}$ is the reference phase at time $t_{0}$; $\nu$, $\dot{\nu}$ and $\ddot{\nu}$ are the pulsar 
rotational frequency and its derivatives. TEMPO attempts to minimize the deviations between the observed 
and model rotational phases using $\chi ^{2}$ minimization.  Timing irregularities are seen as slow, large 
changes in the residuals $\phi$ $-$ $\phi_{m}$, where $\phi$ is the measured 
phase and $\phi_{m}$ is the model phase. 

For a glitch, there is a sudden change in the phase residuals, modeled by an abrupt jump in the frequency and 
its derivative, followed by a relaxation process. The frequency perturbation due to a glitch can be described as 
\citep{Yuan10}
\begin{equation}
\Delta \nu(t) = \Delta \nu_{p} + \Delta \dot{\nu_{p}}t + \Delta \nu_{d} e^{-t/\tau_{d}}
\end{equation}
\begin{equation}
\Delta \dot \nu(t) = \Delta \dot{\nu_{p}} + \Delta \dot{\nu_{d}}e^{-t/\tau_{d}}
\end{equation}
where $\Delta \nu$ and $\Delta \dot \nu$ are the changes in the pulse frequency and its derivative, relative to the
pre-glitch model; $\Delta \nu_{p}$ and $\Delta \dot{\nu_{p}}$ are the persistent change in rotational frequency and its
derivative; $\Delta \nu_{d}$ is the amplitude of the exponentially decaying part of the jump in rotational frequency 
(and $\Delta \dot{\nu_{d}}$ is the corresponding value for the frequency derivative), with a relaxation time 
constant $\tau_{d}$. The total frequency change at the time of the glitch is then given by
\begin{equation}
\Delta \nu_{g} = \Delta \nu_{p} + \Delta \nu_{d}
\end{equation}
The instantaneous change in $\dot \nu$ at the glitch is given by,
\begin{equation}
\Delta \dot{\nu_{g}} = \Delta \dot{\nu_{p}} + \Delta \dot{\nu_{d}}
\end{equation}

\section {Results}
\begin{figure}
\begin{center}
\includegraphics[angle=270,width=0.45\textwidth]{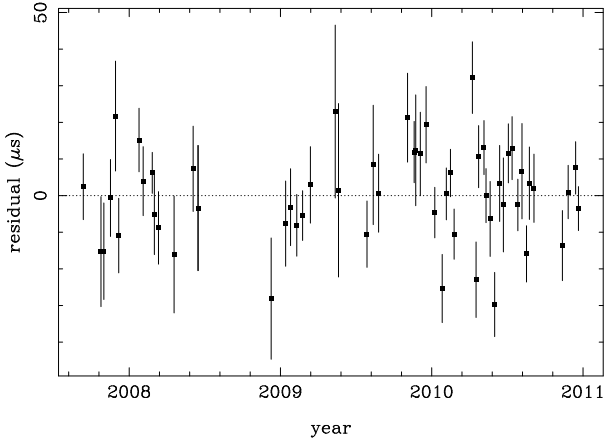}
\caption[Timing residual B1855$+$09]{Timing residuals for the control MSP B1855$+$09, which
show a rms $\sim$ 15 $\mu$s.}
\label{b1855+09_residual}
\vspace{0.1cm} 
%\hrule height 1.5pt
\end{center}
\end{figure}

We first discuss the timing results for the control pulsar and then present the results 
from the timing analysis of the target pulsar J1833$-$1034, using GMRT data spanning 30th 
July 2005 to 11th Jan 2011. The post-fit residuals obtained from the phase connected timing
solution for the control pulsar B1855$+$09 (shown in Fig. \ref{b1855+09_residual}) over the full data 
span of 52 epochs, yield a root-mean-square (rms) value of 15 $\mu$s. This is more than the theoretical, 
expected value ($\sigma_{th}$) of 3 $\mu$s, which is based on the expected S/N of 177 given in Table 1, 
and using $\sigma_{TOA}$ $\simeq$ $\frac{W}{S/N}$ (where $W$ is the pulse width and $S/N$ is the 
{\it expected} signal-to-noise ratio of the average profiles). However, it matches well with the value 
expected for the {\it achieved} S/N (which has a maximum value of 35 in all the observing epochs). 
Nevertheless, our final 610 MHz timing residuals for PSR B1855$+$09 are worse off by factor of 5 
in rms from the 1420 MHz results obtained by \cite{Hobbs06}. In addition to signal to noise limitations, 
there can be effects of interstellar weather that can reduce the accuracy of the TOAs : the pulse arrival 
time at each epoch can have extra deviations due to propagation effects in the interstellar medium, 
which will be larger at the lower frequency.   These results verify, to first order, the proper
working of the pulsar timing set-up at the GMRT. 

\begin{figure}
\begin{center}
\includegraphics[angle=270,width=0.45\textwidth]{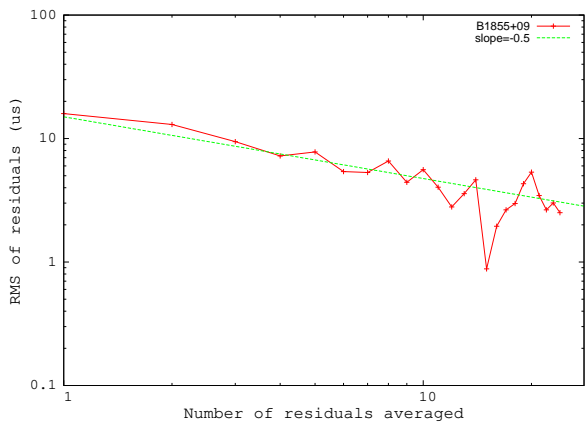}
\caption[Improvement of residual rms of B1855$+$09 with averaging]{RMS of residuals versus number of consecutive residuals averaged, for
the PSR B1855$+$09. Green dashed lines indicate the expected slope of $-$0.5 for uncorrelated residuals.}
\label{b1855+09_check}
\vspace{0.1cm}
%\hrule height 1.5pt
\end{center}
\end{figure}

In order to check for any low-level systematic effects in the timing residuals for this control pulsar,
we investigated the changes in the rms value when adjacent residuals are averaged. For truly white-noise
residuals, this rms should decrease as the square root of the number of residuals averaged.
Fig. \ref{b1855+09_check} shows the results for this on a log-log plot, where the data points are
found to match quite well with a slope of $-$0.5 (green dashed line).  The over-all post-fit residuals
thus exhibit a white-noise behaviour and are likely free from any systematics.
The averaging of post-fit residuals over 166 days (10 TOAs) achieves a rms of 3 $\mu$s
for B1855$+$09, which implies a long-term timing stability of 1 part in 4$\times10^{12}$.
The above results for the control pulsar establish the basic fidelity of the timing pipeline for the
GMRT. 

\begin{figure}
\begin{center}
\includegraphics[angle=270,width=0.45\textwidth]{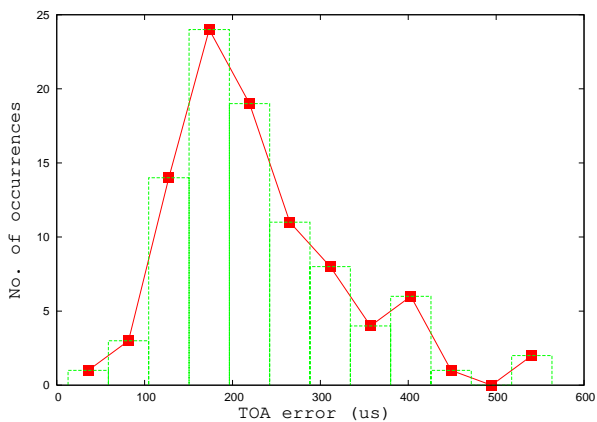}
\caption[TOA error distribution of J1833$-$1034]{Distribution of TOA errors for PSR J1833$-$1034.}
\label{toa_error}
\vspace{0.1cm}
\end{center}
\end{figure}

For PSR J1833$-$1034, since the S/N is typically significantly lower than for the control pulsar, 
to check the data quality for timing purposes we show the distribution of TOA errors in 
Fig. \ref{toa_error}.  The bulk of the values are clustered in the range of 100 to 300 $\mu$s, 
with a small tail of larger values. This skew in the distribution is due to degradation 
of S/N at some epochs, possibly due to fading caused by interstellar scintillation. However, 
the achieved TOA uncertainties are still good enough to detect changes in the residuals of the 
order of several milliseconds due to the occurrence of glitches.  

%%%%%%%%%%%%%%%%%%%%%%%%%%%%%%%%%%%%%%%%%%%%%%%%%%% 
%%%%%%%%%%%%%%%%%%%%%%%%%%%%%%%%%%%%%%%%%%%%%%%%%%%% 
%%%%%%%%%%%%%%%%%%%%%%%%%%%%%%%%%%%%%%%%%%%%%%%%%%%%
%%%%%%%%%%%%%%%%%%%%%%%%%%%%%%%%%%%%%%%%%%%%%%%%%%%%%
\begin{figure}
\begin{center}
\includegraphics[angle=270,width=0.45\textwidth]{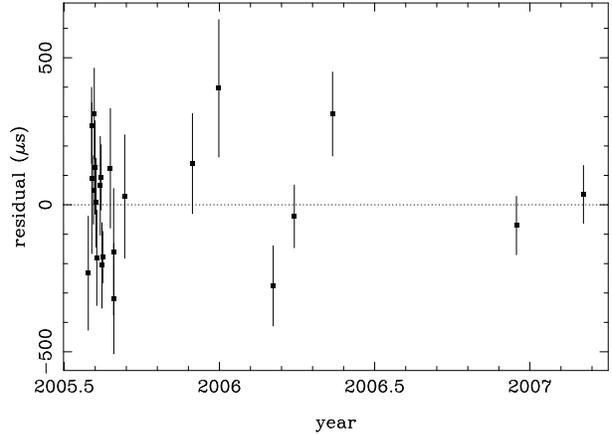}
\caption[Timing residuals of J1833$-$1034 from initial 1.5 years of timing data]
{Timing residuals for the target pulsar PSR J1833$-$1034 from the first 1.5 years of data, 
prior to the first detected glitch at $\sim$ 2007.2.}
\label{J1833-1034_pre_glitch}
\vspace{0.1cm}
\end{center}
\end{figure}
%%%%%%%%%%%%%%%%%%%%%%%%%%%%%%%%%%%%%%%%%%%%%%%%%%%%%

Starting with the initial timing observations for PSR J1833$-$1034, we are able to build up a
phase-connected timing solution (shown in Fig. \ref{J1833-1034_pre_glitch}), till the epoch of 2007.2.
From this 1.5 year data span, we obtain a fairly good timing model for this pulsar,
including a second frequency derivative (see the first row of Table \ref{rotational_param}), and the 
rms of the residuals is around 174 $\mu$s. The reference epoch (MJD) for these measurements is set to the 
epoch which is mid-point of our full data span (i.e. MJD of 54575), for better comparison with the
later models. The pulsar position used in the timing model is the one determined from the Chandra 
observations \citep{Camilo06}. The position derived from our timing solution of 1.5 years of 
phase-connected residuals is within the 3$\sigma$ error bars of this X-ray position. We derive a 
braking index ($n$ $=$ $\nu \ddot\nu$/$\dot\nu^{2}$ ) of 2.168(8) for this pulsar from this initial data span (see last column of 
Table \ref{rotational_param}). 
%%%%%%%%%%%%%%%%%%%%%%%%%%%%%%%%%%%%%%%%%%%%%%%%%%%%
\begin{figure}
\begin{center}
\includegraphics[angle=0,width=0.5\textwidth]{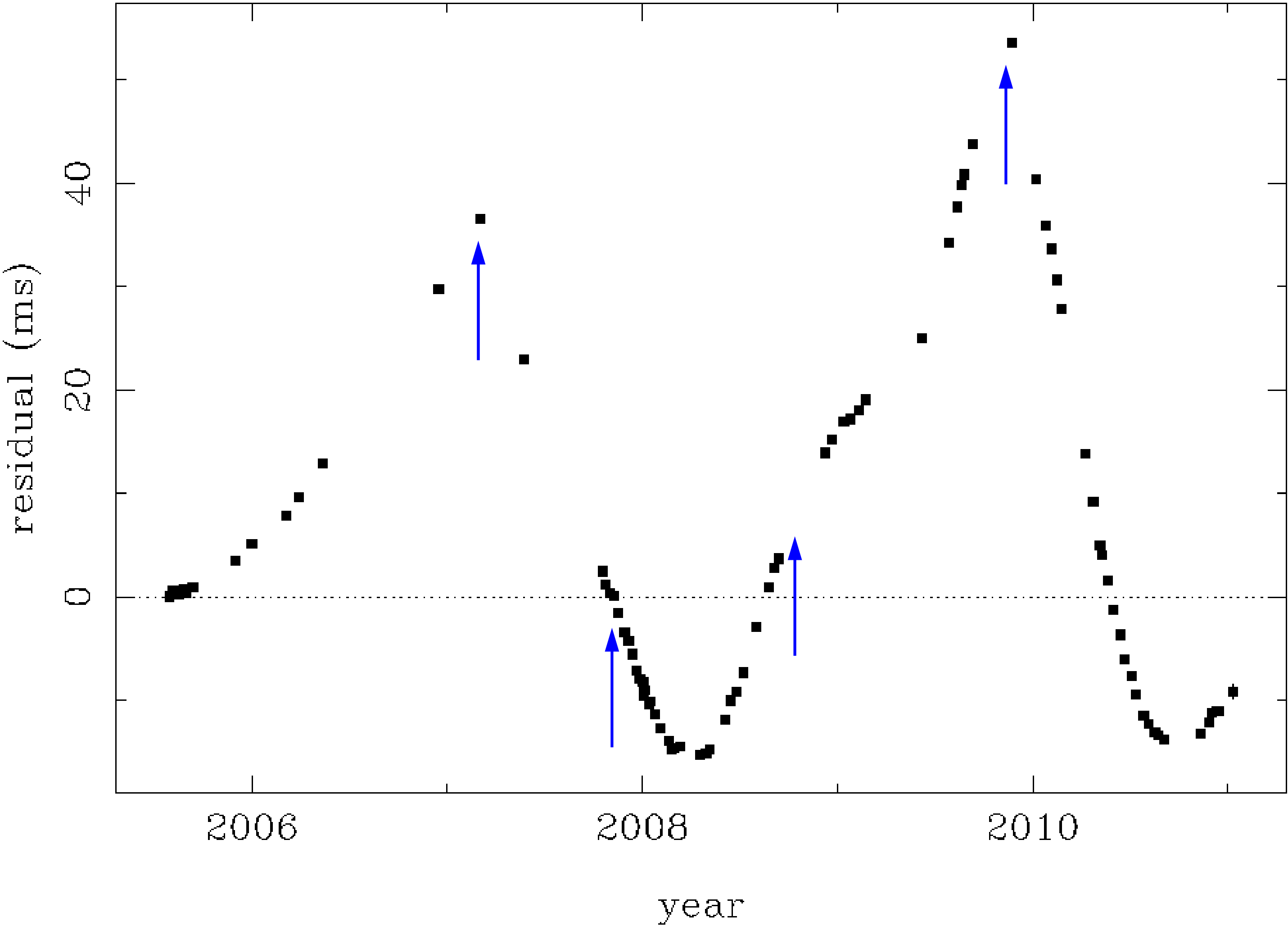}
\caption[Timing residuals of J1833$-$1034 without any glitch models]{Residuals from the full 5.5 yrs of data without modeling 
of any glitch event, showing strong signature of timing noise, as well as evidence for glitches $-$ these are typically 
seen as sudden negative change in the slope of the residuals, and the suspected locations are indicated by arrows. The 
detection and modeling of these glitches are explained in detail in the text, and illustrated in Figs \ref{J1833-1034_glitch0},
\ref{J1833-1034_glitch1}, \ref{J1833-1034_glitch2} \& \ref{J1833-1034_glitch3}.}
\label{J1833-1034_TN_preglitch}
\vspace{0.1cm}
%\hrule height 1.5pt
\end{center}
\end{figure}
%%%%%%%%%%%%%%%%%%%%%%%%%%%%%%%%%%%%%%%%%%%%%%%%%%%%%
\begin{figure}
\begin{center}  
\includegraphics[angle=0,width=0.40\textwidth]{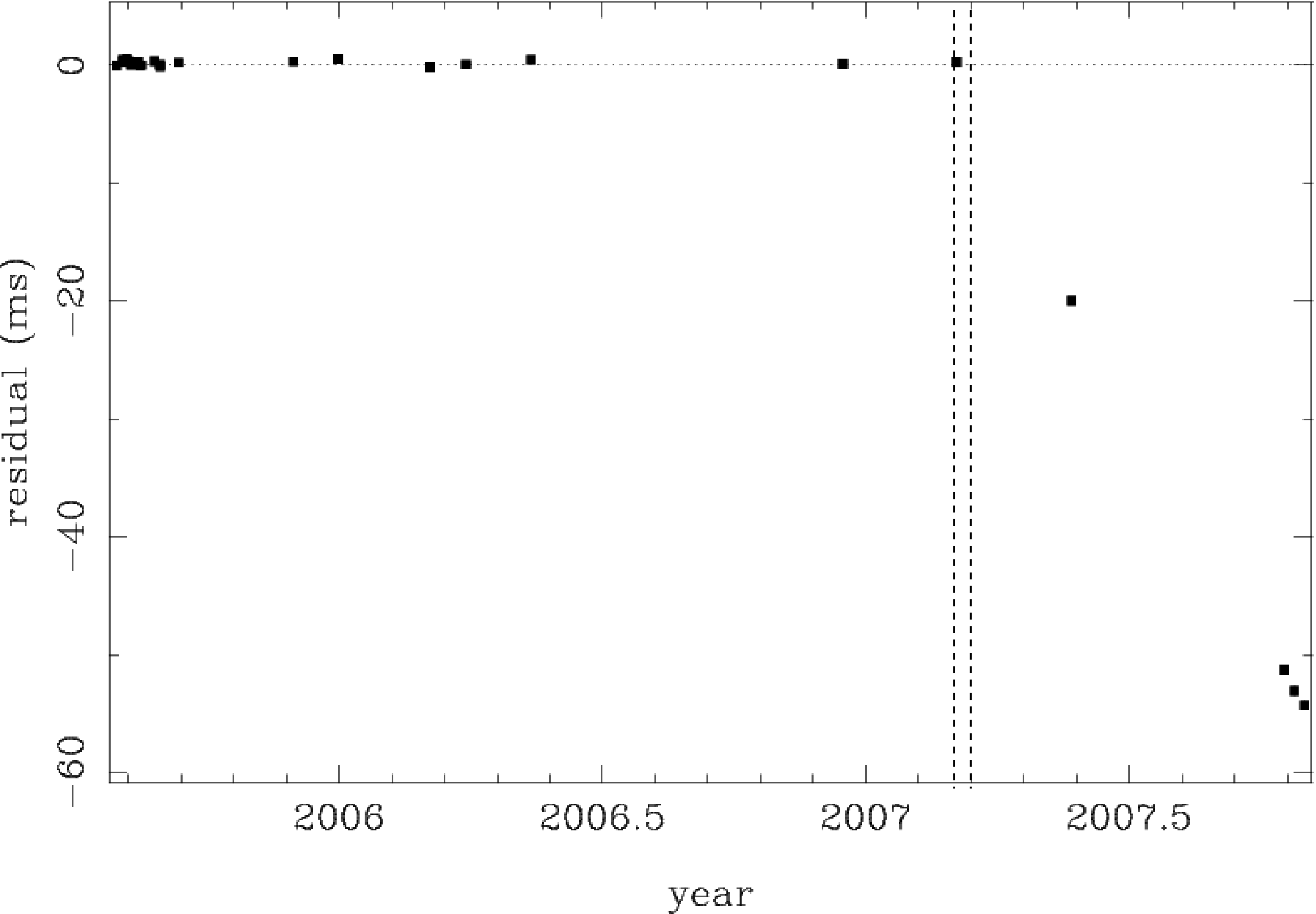}  
\caption[First glitch in PSR J1833$-$1034]{Timing residuals illustrating the first glitch event at $\sim$ 2007.2.
This initial model is obtained from fitting $\nu$, $\dot\nu$ and $\ddot\nu$ to TOAs from 2005.5 up to the suspected 
epoch of the glitch.  The final model yields a glitch with $\Delta\nu_{g}$/ $\nu$ of $3.34\times10^{-9}$, localised 
in time to the interval marked by the dotted lines.}
\label{J1833-1034_glitch0}
\includegraphics[angle=0,width=0.40\textwidth]{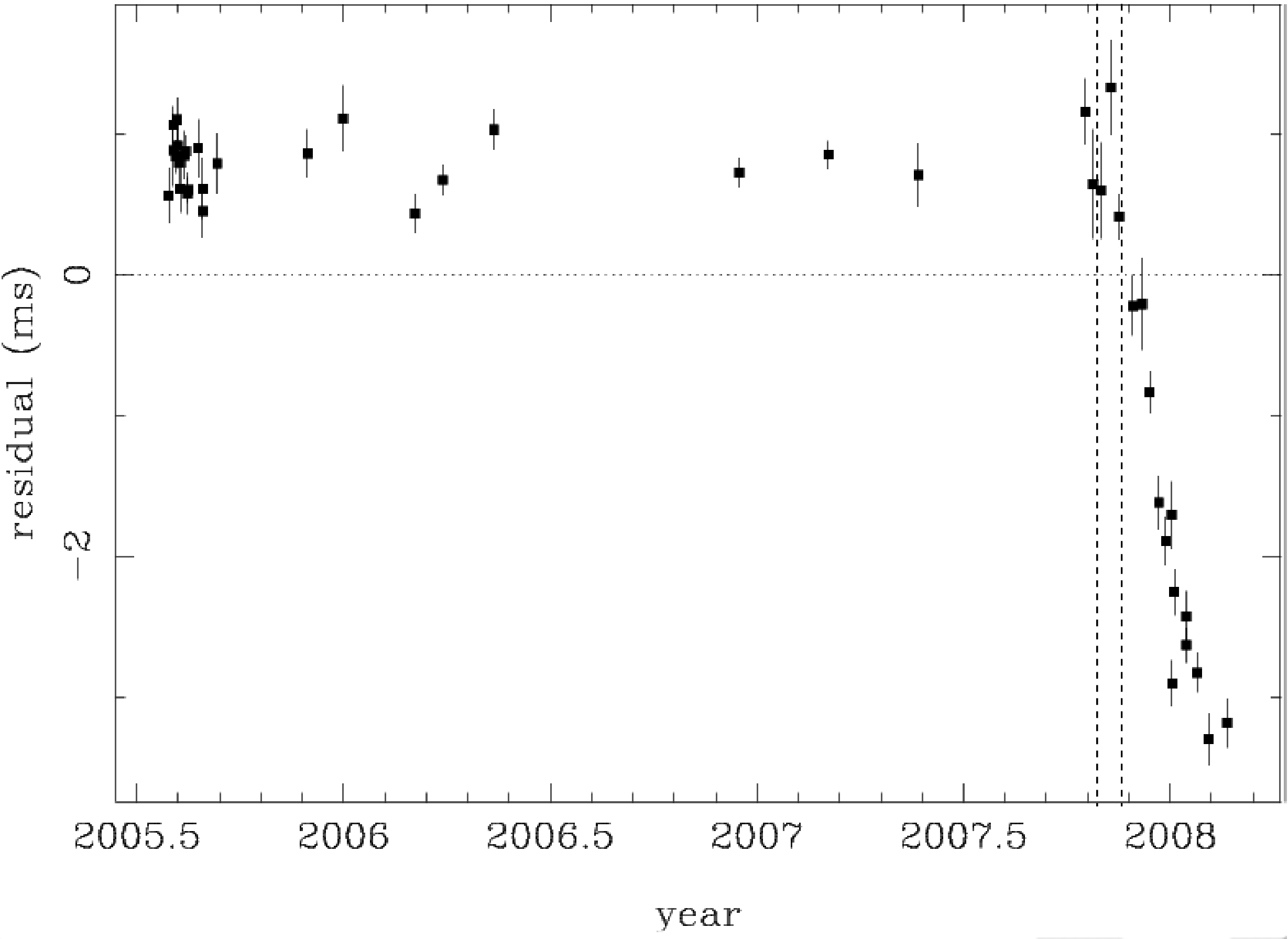}
\caption[Second glitch in PSR J1833$-$1034]{Timing residuals illustrating the second glitch event at $\sim$ 2007.9.
This initial model is obtained from fitting $\nu$, $\dot\nu$ and $\ddot\nu$ and the glitch model for the first 
glitch, to TOAs from 2005.5 up to the suspected epoch of the glitch.  The final model yields a glitch with 
$\Delta\nu_{g}$/ $\nu$ of $1.00\times10^{-9}$, localised in time to the interval marked by the dotted lines.}
\label{J1833-1034_glitch1}
\vspace{0.1cm}
\end{center} 
\end{figure}

\begin{figure}
\begin{center}
\includegraphics[angle=0,width=0.40\textwidth]{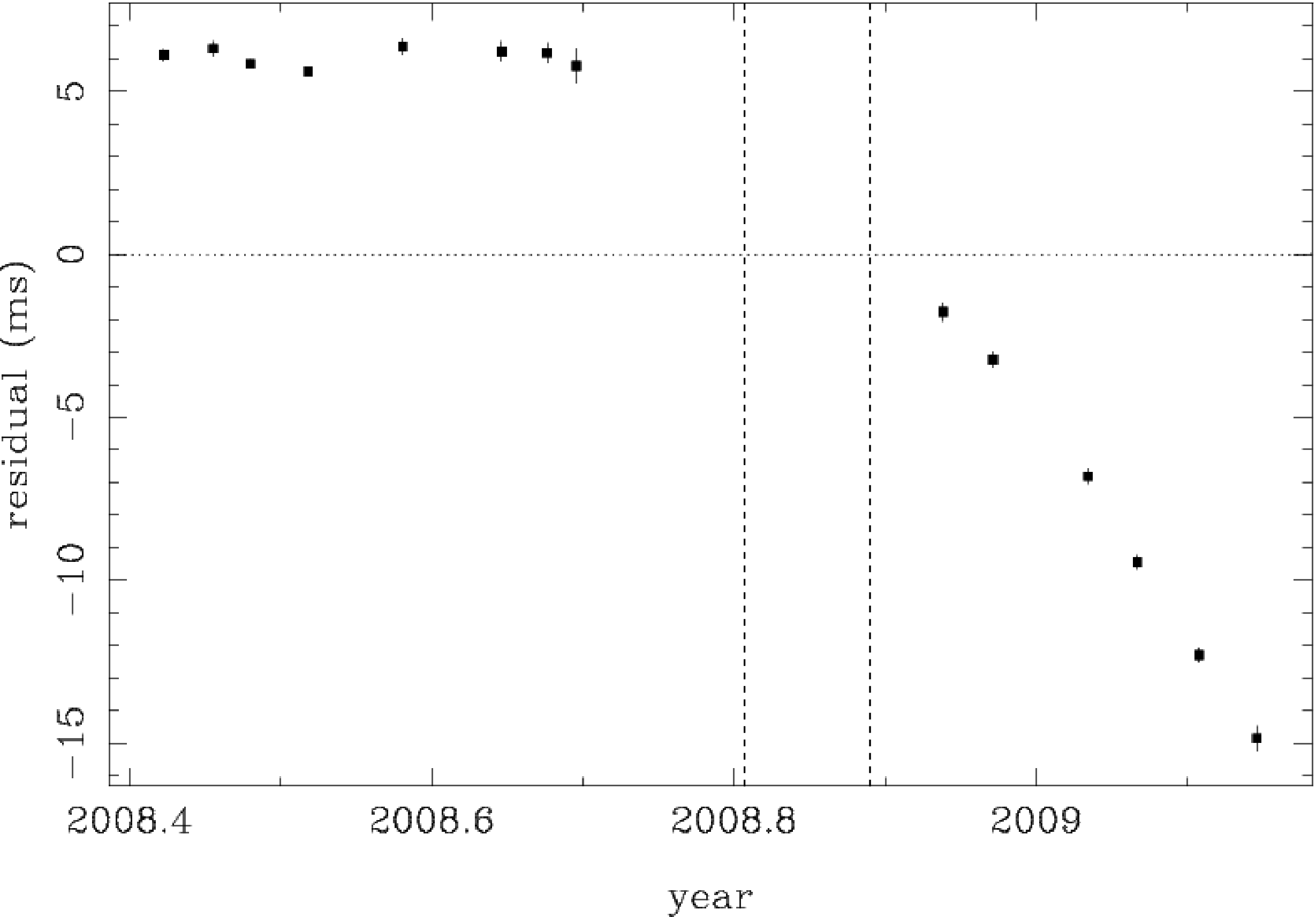}
\caption[Third glitch in PSR J1833$-$1034]{Timing residuals illustrating the third glitch event at $\sim$ 2008.8.
This initial model is obtained from fitting $\nu$ \& $\dot\nu$ over the first 130 days shown here, with $\ddot\nu$ 
held constant at the value obtained from the fit to the first 1.5 years of data.  The final model yields a glitch
with $\Delta\nu_{g}$/ $\nu$ of $1.6\times10^{-9}$, localised in time to the interval marked by the dotted lines.}
\label{J1833-1034_glitch2}
\includegraphics[angle=0,width=0.40\textwidth]{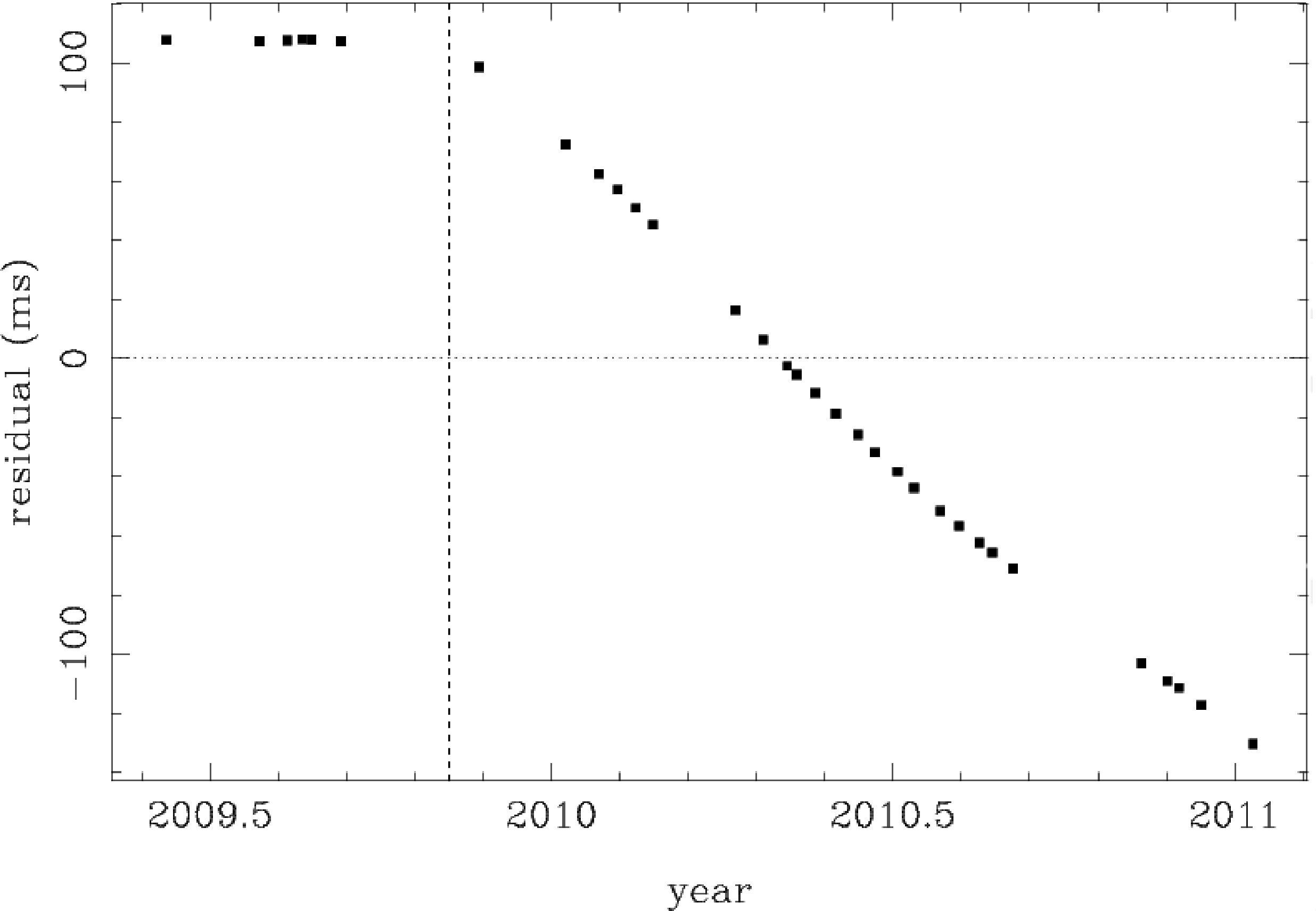}
\caption[Fourth glitch in PSR J1833$-$1034]{Timing residuals illustrating the fourth glitch event at $\sim$ 2009.9.
This initial model is obtained from fitting $\nu$ \& $\dot\nu$ over the first 152 days shown here, with $\ddot\nu$ 
held constant at the value obtained from the fit to the first 1.5 years of data.  The final model yields a glitch
with $\Delta\nu_{g}$/ $\nu$ of $6.9\times10^{-9}$, localised in time to the interval marked by the dotted lines.}
\label{J1833-1034_glitch3}
\vspace{0.1cm}
\end{center}
\end{figure}
%%%%%%%%%%%%%%%%%%%%%%%%%%%%%%%%%%%%%%%%%%%%%%%%%%%%
\begin{table*}
\begin{center}
\caption{Rotational parameters for PSR J1833$-$1034 from different timing solutions. 
The first row is for the timing solution from the initial 1.5 years of data, 
before the first detected glitch. The second row is for the full 5.5 yrs of 
data, without inclusion of any glitch models. The third row is for the full 5.5 
yrs of data, with glitch models fixed to the values derived from the piecemeal
modeling of the glitches. The last row is for the solution from the final global 
fit, including all four glitches (2 free parameters each, with glitch epochs fixed 
to the values obtained from the piecemeal fittings), and four frequency 
derivatives. All errors are 1$\sigma$ values.}
\vspace{0.3cm}
\label{rotational_param}
\begin{tabular}{|l|c|c|c|c|c|c|c|c|c|c|c|c|c|c|c|c}
\hline
No of glitches&Ref Epoch &Data span 	 &No. of&$\nu$      	 &$\dot{\nu}$        &$\ddot{\nu}$       &Residual &Braking \\
fitted    &(MJD )  &MJD		 &TOAs	&($s^{-1}$) 	 &($10^{-11}s^{-2}$) &($10^{-22}s^{-3}$) &(ms) &Index   \\\hline
          &        &		 &	&           	 &                   &                   &         &       \\
$-$	  &54575   &53581$-$54164&22    &16.159357125(2)&$-$5.275017(9)       &3.73(1)	         &0.174    &2.168(8)\\
$-$	  &54575   &53581$-$55572&94    &16.15935713057(2)&$-$5.27507199(3)   &3.6006(2)	 &15.4     &2.0891(1)\\
$4$       &54575   &53581$-$55572&94	&16.15935711448(2)&$-$5.27507291(3)   &3.6232(2)         &2.20     &2.1041(1) \\
$4$       &54575   &53581$-$55572&94	&16.15935711336(3)&$-$5.2751130(1)    &3.197(1)          &0.512    &1.8569(6) \\\hline
\end{tabular}
\vspace{0.3cm}\\
\end{center}
\end{table*}
%%%%%%%%%%%%%%%%%%%%%%%%%%%%%%%%%%%%%%%%%%%%%%%%%%%%

Fig. \ref{J1833-1034_TN_preglitch} shows the timing residuals for the full data set (94 epochs
spanning 5.5 years) for this pulsar, relative to a simple slow-down model including the pulsar 
spin frequency and its first two derivatives. The best fit model parameters from this are listed in the second row 
of Table \ref{rotational_param}.  The residuals, with a relatively large rms of 15.4 ms, are clearly 
dominated by non-random, low frequency timing noise effects. The amplitude of this timing 
noise is a strongly increasing function of the length of the data span. The effect of  this 
timing noise was probably not detected for the initial data span of 1.5 years, as the fitting of the 
spin-frequency and its two derivatives can mask most of the low-frequency trends. In these timing 
noise dominated residuals of Fig. \ref{J1833-1034_TN_preglitch}, the presence of glitches can be 
distinguished by sudden changes in the slope of the curve.  Clear events are seen at 2007.2 and 
2009.9, and less likely ones at 2007.9 and 2008.8, all of which are marked by arrows in 
Fig. \ref{J1833-1034_TN_preglitch}.

The presence of a glitch is confirmed by taking relatively shorter stretches of data around the 
suspected glitch event, and doing a local timing fit to the TOAs, starting with a model for the
data prior to the glitch.  Sudden, systematic deviation of the residuals from a smooth behaviour 
is taken as the signature of the occurrence of a glitch (as seen in Figs \ref{J1833-1034_glitch0}, 
\ref{J1833-1034_glitch1}, \ref{J1833-1034_glitch2} \& \ref{J1833-1034_glitch3}).  Detailed 
modeling is then carried out to estimate the glitch epoch, and the changes in frequency and 
frequency derivative at the glitch.  The best possible value of the glitch epoch is estimated 
by minimising the phase increment required to obtain a phase-connected solution over the interval
around the glitch epoch \citep{Janssen06}. The measurement uncertainty of the glitch epoch is 
obtained from the corresponding 3$\sigma$ limit of the glitch phase increment parameter.

Starting with an initial timing model having $\nu$, $\dot{\nu}$ and $\ddot{\nu}$, we find 
that the first glitch (Fig. \ref{J1833-1034_glitch0}) occurred at MJD $=$ 51469 ($\pm$ 7), 
with a fractional change in the rotational frequency (${\Delta \nu_{g}} / {\nu}$) of 
3.34$\times 10^{-9}$. The modeling for this glitch includes 26 TOAs observed over 824 
calender days, of which there are 236 days of data after the glitch event. 
Fig. \ref{J1833-1034_glitch1} shows the second glitch event, which is best fit by a 
fractional increase in rotational frequency of 1.00$\times 10^{-9}$ at MJD $=$ 54423 ($\pm$ 9). 
As we have a fairly good timing model for the first 1.5 years (including the first glitch event), 
without any timing noise effects, the pre-glitch interval for this second glitch includes all 
of these TOAs over 842 calender days, with a pre-fit model of $\nu$, $\dot{\nu}$, $\ddot{\nu}$ 
and the derived parameters of the first glitch.  There is a third glitch (Fig. \ref{J1833-1034_glitch2})
detected at MJD $=$ 54750 ($\pm$ 15) with a fractional change in the rotational frequency of 
1.6$\times 10^{-9}$.  The last glitch event (Fig. \ref{J1833-1034_glitch3}) observed at 
MJD $=$ 55142 ($\pm$ 2) yields a fractional change in the rotational frequency of 
6.9$\times 10^{-9}$.  In order to minimse the effects of timing noise, the pre-glitch interval 
for the third glitch includes TOAs over 130 days and the fourth glitch includes TOAs over 152 
days, with the pre-fit model having $\nu$, $\dot{\nu}$ \& $\ddot{\nu}$.  But since the pre-fit 
data span over smaller intervals, the fit uses $\nu$, $\dot{\nu}$ as free parameters, with 
$\ddot{\nu}$ kept constant to the value derived from the initial 1.5 yrs of data.
Inclusion of TOAs over larger spans increases the influence of timing noise, where the 
residuals depart from the simple spin-down model with $\nu$, $\dot\nu$ and $\ddot\nu$,
making it harder to detect the glitches accurately. 
There are also small changes 
in slow-down rate, of the order of $10^{-5}$, observed at the glitch epochs. The new timing 
models, after inclusion of the glitch parameters, yield phase-connected timing residuals with 
rms values of 177 $\mu$s, 216 $\mu$s, 227 $\mu$s and 1.5 $m$s respectively, for the four 
cases. 

\begin{figure}
\begin{center}
\includegraphics[angle=270,width=0.45\textwidth]{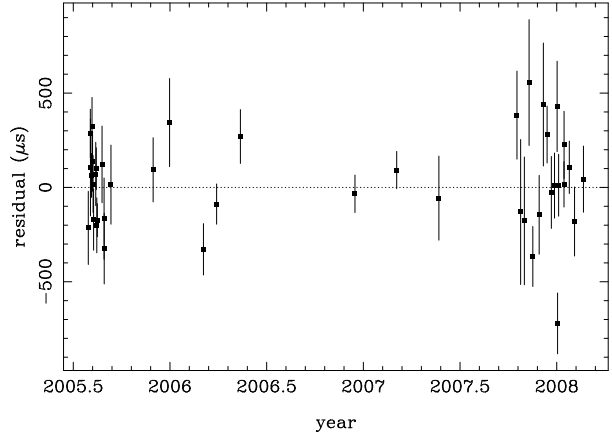}
\caption[Post-fit residuals for first and second glitch]{Post-fit residuals after fitting 
for the first two glitches, for TOAs spanning the first 936 days.  The rms is 216 $\mu$s.}
\label{J1833-1034_glitch01_post-fit}
\vspace{0.1cm}
\end{center}
\end{figure}

\begin{figure} 
\begin{center}
\includegraphics[angle=270,width=0.45\textwidth]{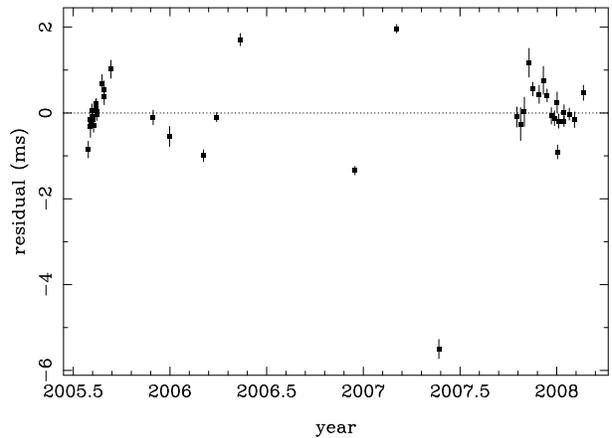}
\caption[Post-fit residuals without inclusion of first and second glitch]{Post-fit residuals 
(for TOAs spanning the first 936 days) after fitting with a model having $\nu$ and 8 frequency 
derivatives, without inclusion of any glitch models, which amounts to 9 free parameters, as 
for the model with glitches. The rms is 951 $\mu$s, and there are large deviations at the
epoch of the first glitch.}
\label{J1833-1034_glitch01_high-deri}
\vspace{0.1cm}
\end{center}
\end{figure}

It is sometimes possible that, for data that are relatively sparsely sampled and
have significant amount of timing noise (both of which are somewhat true for the present case), 
there can be large deviations in residuals with respect to the basic spin-down model
of $\nu$, $\dot{\nu}$ \& $\ddot{\nu}$, which $\it may$ mimic glitch-like behaviour.  In order 
to discriminate between the effect of timing noise and genuine glitches, we investigated 
the pre-fit and post-fit residuals around the glitch epochs by fitting with higher frequency 
derivatives without inclusion of any glitch model.  For example, for the case of TOAs 
spanning over the first 936 days (including glitch-1 and glitch-2), the rms for the 
post-fit residuals is 216 $\mu$s (shown in Fig. \ref{J1833-1034_glitch01_post-fit}). 
The model includes two glitches and a spin-down model with  $\nu$, $\dot{\nu}$ \& $\ddot{\nu}$, 
which amount to 9 free parameters. The same span of TOAs can also be fitted with a
model having $\nu$ and 8 frequency derivatives, without inclusion of any glitch 
models, which also amounts to 9 free parameters, as for the model with glitches.
The post-fit residuals shown in Fig. \ref{J1833-1034_glitch01_high-deri}, 
have a much larger rms value of 951 $\mu$s and also show large discontinuities, including
at the glitch epoch.  Similar effects were found for the data around the 3rd and 4th
glitches.  This illustrates the fact that some of the large TOA variations that we see
for this pulsar, over and above the basic spin-down model, can not be satisfactorily 
explained with a model of timing noise characterised by higher order derivatives, 
but are better modeled with discrete glitch events.

%%%%%%%%%%%%%%%%%%%%%%%%%%%%%%%%%%%%%%%%%%%%%%%%%%%%%
%%%%%%%%%%%%%%%%%%%%%%%%%%%%%%%%%%%%%%%%%%%%%%%%%%%%
%%%%%%%%%%%%%%%%%%%%%%%%%%%%%%%%%%%%%%%%%%%%%%%%%%%%

\begin{figure}
\begin{center}
\includegraphics[angle=0,width=0.45\textwidth]{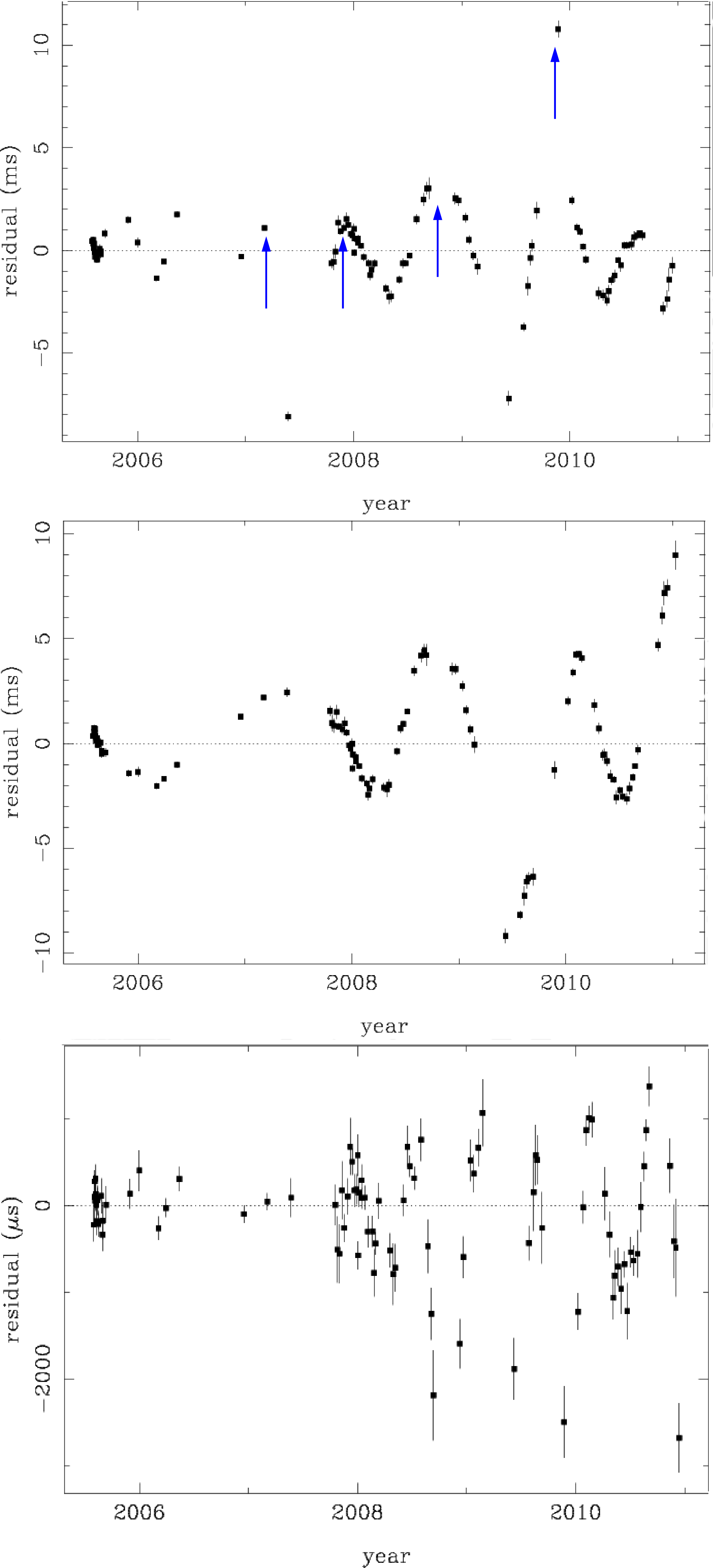}
\caption[Post-fit residuals for PSR J1833$-$1034]{Timing residuals for PSR J1833$-$1034, for different types
of global fits, all having the same number of 13 free parameters. The top panel shows residuals with 
12 frequency derivatives fitted, without any glitch models. 
The suspected glitch epochs are marked by arrows. At the first, third and fourth glitch epochs there is clear
evidence for discontinuities in the residuals. The middle panel shows residuals with a model of four glitches, 
derived from piecemeal fitting and two derivatives of pulsar spin-frequency.  The presence of low-frequency 
timing noise is clearly seen in these residuals, which otherwise show a smooth behaviour.  The bottom panel 
shows residuals obtained from a global fit of four glitch models and four frequency derivatives. The rms is
significantly lower and the residuals are much more whitened in nature, indicating that this model gives
the best global timing fit to this data set.}
\label{glitch0123.postfit}
\vspace{0.1cm}
%\hrule height 1.5pt
\end{center}
\end{figure}

Finally, to obtain a global model for the entire data span, we have compared the 
following three approaches :
(i) taking the models for the 4 glitches obtained from the piecemeal fits to the 
individual glitches as fixed and then fitting for $\nu$, $\dot{\nu}$ \& $\ddot{\nu}$
over the entire data span;  (ii) a fit to the full data span using $\nu$ and up to 12 frequency
derivatives (the maximum allowed by TEMPO) without any glitch models included; and
(iii) a global fit for 4 glitches, $\nu$ and first four frequency derivatives, 
to achieve the same count of 13 free parameters as in case (ii) $-$ each
glitch contributes 2 free parameters (spin-frequency increment and change 
in spin-down rate), as the glitch epochs are fixed to the values obtained 
during the piecemeal fittings, by minimising the phase increment at the 
glitch epoch. Case (i) gives the residuals shown in the middle panel of Fig. \ref{glitch0123.postfit},
with an rms of 2.2 ms, and fairly smooth behaviour with large swings, typical of timing
noise.  Results from this fit are given in the third row of Table \ref{rotational_param}.
Case (ii) gives the residuals shown in the top panel of Fig. \ref{glitch0123.postfit}.
Though the rms is 1.4 ms, the variations of the residuals show sudden, large jumps 
(as at the epoch of the first glitch) and also sharp, cuspy variations (as at the epoch
of the 4th glitch).  For case (iii), global fits with all 4 glitches (using the results
from the piecemeal fits as the starting pre-fit model) and increasing number
of frequency derivatives were tried, and the following was found : for 4 glitches plus 
$\nu$, $\dot{\nu}$ \& $\ddot{\nu}$, the residuals are 1.2 ms and the behaviour is 
qualitatively similar to case (i); 
for the case of two more derivatives added to achieve the same number of 13 degrees of freedom as it was 
for the case (ii), the residuals (shown in the bottom panel of Fig. \ref{glitch0123.postfit}) reduce significantly 
to 0.5 ms and also the slow, large fluctuations typical of timing noise are noticeably
suppressed.  The glitch parameters are not very different from those obtained
from the piecemeal fits.  The results from this model are summarised in the 4th row
of Table \ref{rotational_param} and the final glitch parameters are given in 
Table \ref {glitch-summary}.

From the above, we argue that the best timing model is that given by a global fit of 4 
glitches and 5 spin frequency terms, which gives the best global fit to the data and reduces
the residuals to a minimum.  The attempt to fit the TOAs with a pure 
timing noise model having large number of derivatives does not give acceptable results :
both for localised fits to data sets that span individual glitches, as well as for the
global data set.  For such cases, the rms of the residuals is larger and/or the residuals 
show uncharacteristically large, abrupt changes.  We take the results from this fit as 
the final timing model for this data set.  These results are summarised in the last row
of Table \ref{rotational_param} and in Table \ref {glitch-summary}.

Now in order to measure the amount of timing noise present in this pulsar, we 
have used the definition given by \cite{Arzoumanian94},
\begin{equation}
\Delta_{8} = log_{10}(\frac{1}{6\nu}|\ddot{\nu}|t^{3})
\end{equation}
where the spin-frequency, $\nu$ and its second derivatives, $\ddot{\nu}$, are measured over a $t$ $=$ $10^{8}$ s 
interval. We have used first 3.16 years of data for PSR J1833$-$1034 to estimate the value of $\Delta_{8}$ as 
0.5, which follows the correlation between timing noise and spin-down rate, i.e. the younger pulsars with 
larger spin-down rate exhibit more timing noise than older pulsars, seen by \cite{Arzoumanian94} and later 
by \cite{Hobbs10}.  

The value of the braking index determined from the final global fit is 1.8569(6).
This braking index is much less than 3, which is in general agreement 
with the values obtained for other young pulsars having reliable estimates 
for this quantity. For example, for Crab pulsar,
$n$ $=$ 2.509(1) \citep{Lyne88,Lyne93}, for PSR J1846$-$0258, $n$ $=$ 2.65(1)
\citep{Livingstone06}, for PSR B0540$-$69, $n$ $=$ 2.140(9) \citep{Livingstone05}, for
PSR B1509$-$58, $n$ $=$ 2.837(1) \citep{Kaspi94} and for PSR J1119$-$6127, $n$ $=$ 2.684(2) \citep{Weltevrede11}. 
A value of n $<$ 3 indicates that simple magnetic dipole model does not completely explain 
spin-down evolution of pulsars. Particle outflow in the pulsar wind can also carry away some of its 
rotational kinetic energy.

%%%%%%%%%%%%%%%%%%%%%%%%%%%%%%%%%%%%%%%%%%%%%%%%%%%%%
%%%%%%%%%%%%%%%%%%%%%%%%%%%%%%%%%%%%%%%%%%%%%%%%%%%%
\begin{table*}
\begin{center}
\caption{The parameters of all the four glitches detected in PSR 1833$-$1034, as determined from
a global fit to the timing data. The errors on the least significant digit are at 1$\sigma$ level.}
\label{glitch-summary}
\begin{tabular}{|l|c|c|c|c|c|c|c|c|c|c|c|c|c|c|c|c}
\hline
Glitch epoch   &Date	     &Fit span     &$\frac{\Delta\nu_{g}}{\nu}$  &$\frac{\Delta\dot{\nu}_{g}}{\dot{\nu}}$ \\
(MJD)          &             &(MJD)        &($10^{-9}$)                  & ($10^{-5}$)                     \\\hline
54169 ($\pm$7) &5th Mar2007  &53581$-$54405&3.11(5)                      &1.4(2)                           \\
54423 ($\pm$9) &12th Nov2007 &53581$-$54517&1.09(6)                      &4.0(3)                           \\
54750 ($\pm$15)&11th Nov2008 &54620$-$54885&3.55(6)                      &$-$7.7(2)                        \\
55142 ($\pm$2) &6th Nov2009  &54990$-$55572&7.50(8)                      &$-$9.9(2)                        \\\hline
\end{tabular}
\end{center}
\end{table*}
%%%%%%%%%%%%%%%%%%%%%%%%%%%%%%%%%%%%%%%%%%%%%%%%%%%%

\section {Discussion}
Our timing study of the young pulsar J1833$-$1034 associated with the galactic SNR G21.5$-$0.9 shows
clear evidence of frequent glitches in the pulsar's rotational history.  We find as many as 4 glitches
over the observing span of 5.5 years.  Compared to the typical range of glitch amplitudes mentioned in
section 1, the fractional changes in the rotational frequency seen for this pulsar are relatively small,
ranging from $1\times10^{-9}$ to $7\times10^{-9}$.  This behaviour is similar to the Crab pulsar, which
shows $\Delta \nu / \nu$ $\sim$ $10^{-8}$; whereas the Vela pulsar exhibits larger glitches, generally with
$\Delta \nu / \nu$ $>$ $10^{-6}$. As the amplitude of a glitch is related to the amount of stress built up in 
the pinned vortices, one might expect some correlation between the amplitude of glitches and the inter-glitch 
interval.  Pulsars that have small amplitude glitches do tend to show smaller interval between glitches 
(as observed for PSR J0537$-$6910 by \cite{Middleditch06} and for PSR B1642$-$03 by \cite{Shabanova09}), 
and this is borne out in the case of PSR J1833$-$1034 as well.  Clearly, this pulsar falls under the 
category of pulsars that exhibit relatively frequent, but low amplitude glitches.  

%%%%%%%%%%%%%%%%%%%%%%%%%%%%%%%%%%%%%%%%%%%%%%%%%%% 
\begin{figure}
\begin{center}
\includegraphics[angle=270,width=0.45\textwidth]{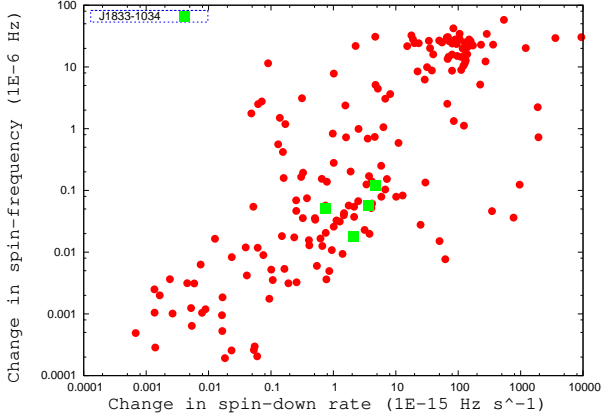}
\caption[Changes in spin-down rate with spin-frequency at the glitch]{Plot of changes in the
pulsar spin-down rate, $|\Delta \dot{\nu}|$, with spin-frequency due to glitches. The data points
(red circles) are from the glitch table of the ATNF pulsar catalog. The data points for PSR 1833$-$1034 
are denoted by squares.}
\label{J1833-1034_glitch_step}
\vspace{0.1cm}
%\hrule height 1.5pt
\end{center}
\end{figure}
%%%%%%%%%%%%%%%%%%%%%%%%%%%%%%%%%%%%%%%%%%%%%%%%%%%

PSR J1833$-$1034 also shows small but permanent changes in the slow-down rate at the glitches, and the
typical fractional change of $\dot{\nu}$ is a few parts in $10^{-5}$ (Table \ref {glitch-summary}). 
These small increases in $|\dot{\nu}|$ are thought to be due to the decrease in the effective moment 
of inertia of the crust, which includes all components of the star dynamically coupled to the crust. 
Decoupling of the superfluid moment of inertia during the unpinning state reduces the entire moment 
of inertia of the star. However, for the third and fourth glitches, we observe a decrease in $\dot\nu$. 
This sign change of $\dot \nu$ may imply a small increase in moment of inertia or a small decrease 
in spin-down torque at the time of the glitch.

We did not detect any exponential recovery or decay with time after the glitches in our data, for either 
the change in rotational frequency or its derivative.  This may imply that there are only permanent 
changes in the rotational parameters when this pulsar glitches.  However, there is also a possibility 
that this may be due to the fact that our sampling  interval for the timing properties of this pulsar $-$ 
about a week to 10 days $-$ is somewhat coarser than what may be required to adequately sample the 
expected decay time-scale for such low amplitude glitches. For example, in case of the Crab pulsar, 
for glitches with an amplitude of the order of $\sim$ $10^{-8}$, the exponential decay time-scale is 
of $\sim$ 10 days \citep{Wong01}.  Such time scales would be hard to detect in our timing data, and 
would need a much more intensive campaign of observations.

For the general pulsar population it is found that glitches with small $\Delta \nu$ also have small
changes in $|\dot{\nu}|$.  This is shown in Fig. \ref{J1833-1034_glitch_step}, using the database of
the glitch table in the ATNF pulsar catalog \footnote{see http://www.atnf.csiro.au/research/pulsar/psrcat/glitchTbl.html}, 
where a clear correlated trend can be seen. Our results of the glitch parameters 
for PSR J1833$-$1034 shows that this pulsar follows this trend quite well.

The level of strength and frequency of occurrence of glitches in a pulsar can be quantified by the
glitch activity parameter, $A_{g}$, defined as the mean change in frequency per unit time
owing to glitches \citep{Lyne99}:
\begin{equation}
A_{g} = \frac{1}{T} \sum {\Delta\nu_{g}}
\end{equation}
where $\sum {\Delta\nu_{g}}$ is the total increase of the frequency owing to all the glitches 
over an interval of $T$. Glitches are considered as events of angular momentum transfer from the 
superfluid interior to the crust of the neutron star. The same rate of angular momentum transfer can be 
achieved with frequent small glitches or occasional larger ones. The glitch activity parameter combines 
the amplitude and frequency of angular momentum loss due to glitches over the interval of $T$. $A_{g}$ is relatively 
insensitive to the additional discovery of smaller glitches as the quality of a given data set improves, 
and hence it can be used as a long-term indicator of glitch effects \citep{Wong01}. We find 
$A_{g}$ $=$ $1.53\times10^{-15}$ $s^{-2}$ for PSR J1833$-$1034.  

\begin{figure}
\begin{center}
\includegraphics[angle=270,width=0.45\textwidth]{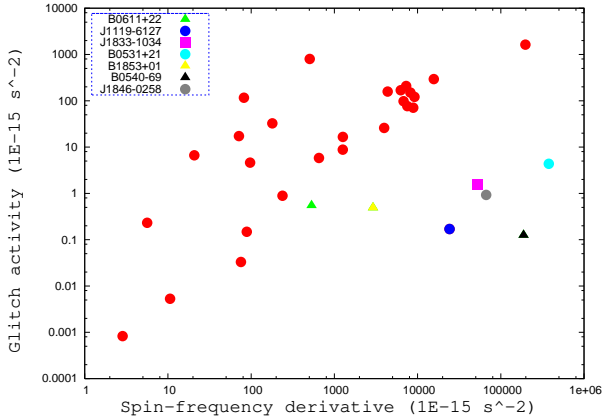}
\caption[Glitch activity of J1833$-$1034 as function of spin-down rate]{Plot of glitch activity parameter, 
$A_{g}$, versus the spin-down rate, $\dot{\nu}$, of pulsars. Circles are data points taken from literature 
and triangles are data points taken from Torun observations. The glitch activity of PSR 1833$-$1034 is 
denoted by the square.} 
\label{J1833-1034_glitch_activity}
\vspace{0.1cm}
%\hrule height 1.5pt
\end{center}
\end{figure}

\begin{figure}
\begin{center}
\includegraphics[angle=270,width=0.45\textwidth]{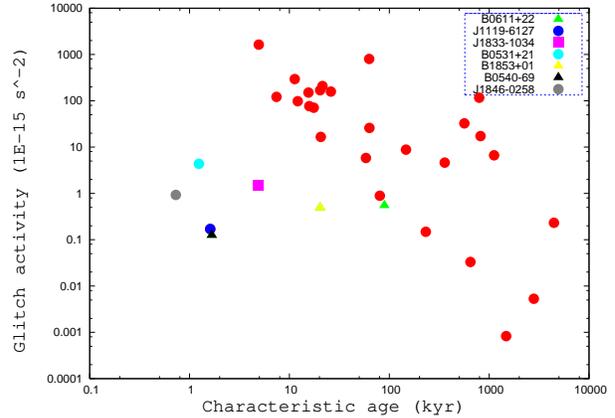}
\caption[Glitch activity of J1833$-$1034 as function of age]{Plot of glitch activity parameter, $A_{g}$, 
versus characteristic age of pulsars. Circles are data points taken from literature and triangles are 
data points taken from Torun observations. The glitch activity of PSR 1833$-$1034 is denoted by the square.}
\label{J1833-1034_glitch_activity_withage}
\vspace{0.1cm}
%\hrule height 1.5pt
\end{center}
\end{figure}

Fig. \ref{J1833-1034_glitch_activity} shows the range of known values of $A_{g}$, as well
as its dependence on $|\dot{\nu}|$, for a collection of 32 pulsars.  The data are mostly from
literature (circles), except for a few points (triangles for B0611$+$22, B1853$+$01 and B0540$-$69) which are from unpublished results from
observations at the Torun Radio Telescope by one of us (Wojciech Lewandowski).  The literature
references are as follow :  \cite{Lyne00} (B0833$-$45, B1325$-$43, B1535$-$56, B1641$-$45, B1727$-$33, B1736$-$29, B1758$-$23,
 B1800$-$21, B1823$-$13, B1830$-$08, B1859$+$07, B2224$+$65, B0355$+$54, B0525$+$21 and B1737$-$30), \cite{Wang00} (B0833$-$45, B1046$-$58, J1105$-$6107,
 J1123$-$6259, B1338$-$62, B1610$-$50, B1706$-$44, B1727$-$47, B1758$-$23, B1757$-$24 and B1800$-$21,), \cite{Wong01} (B0531$+$21), \cite{Hobbs02} 
(J1806$-$2125), \cite{Urama02} (B1737$-$30), \cite{Weltevrede11} (J1119$-$6127), \cite{Shabanova00} (B1822$-$09), \cite{Middleditch06}) (J0537$-$6910), 
\cite{Livingstone06} (J1846$-$0258). 
Though there is some scatter present, for a majority of pulsars 
there is an overall trend of increasing $A_{g}$ with increasing $|\dot{\nu}|$.
This trend is mirrored in a plot of glitch activity versus characteristic age, as shown in  Fig. 
\ref{J1833-1034_glitch_activity_withage} : glitch activity is higher for younger pulsars with 
characteristic age $\sim$ 10 kyr, and as the characteristic age increases, the activity falls off.
These effects could be due to the fact that the flow of the angular momentum from the
interior decreases with age (or increases with $|\dot{\nu}|$). There are a few pulsars with relatively 
higher values of $|\dot{\nu}|$ (or relatively smaller values of characteristic age) that have somewhat 
lower values of $A_{g}$, and hence lie off the main curve.
Detailed investigation shows that these are a group of very young pulsars (i.e. low characteristic age), 
such as the Crab, PSR J1119$-$6127, PSR B1853$+$01, PSR J1846$-$0258 and PSR B0540$-$69. We find that our 
young pulsar J1833$-$1034 fits in very well with this group.

Glitches are thought to be caused by the release of stress built up during the regular spin-down of 
the pulsar. This stress on the pinned vortices in the superfluid interior gets released to the 
solid crust by a collective unpinning of many vortices. This unpinning process results in a sudden spin-up 
of the crust due to this discontinuous transfer of angular momentum from the interior, which in-turn is manifested 
in a change in observed pulsar frequency. Frequent, low amplitude glitches implies that the release of the 
built up stress happens in a more uniform and continuous manner than for pulsars which show few, large amplitude 
glitches.  In other words, for the younger pulsars with larger slow-down rates, the flow of the angular momentum 
from the interior seems to be a smoother process. According to \cite{McKenna90}, the 
higher internal temperature associated with the younger neutron stars might prevent the build up of larger stresses. 
In such cases, stresses on the pinned vortices get relieved by thermal drift of the vortices from one pinning site 
to another in a gradual fashion, resulting in frequent low amplitude glitches. Hence such pulsars may constitute 
a distinct class of glitching pulsars: younger pulsars with lower glitch activity and higher internal temperatures. 
These relatively young pulsars may evolve towards the normal trend (i.e. towards right on 
Fig. \ref{J1833-1034_glitch_activity_withage}) as they age.

%%%%%%%%%%%%%%%%%%%%%%%%%%%%%%%%%%%%%%%%%%%%%%%%%%% 
%%%%%%%%%%%%%%%%%%%%%%%%%%%%%%%%%%%%%%%%%%%%%%%%%%% 
%%%%%%%%%%%%%%%%%%%%%%%%%%%%%%%%%%%%%%%%%%%%%%%%%%%%

\section{Summary and future scope}
In this paper we have presented results for four glitches detected in PSR J1833$-$1034, from 5.5 years
of timing observations at the GMRT. 
These glitches show fractional change of the rotational frequency ranging from  1$\times10^{-9}$ to
7$\times10^{-9}$, with no evidence for any appreciable relaxation of the rotational frequency
after the glitches. The fractional changes observed in the frequency derivative for this
pulsar are of the order of $10^{-5}$. This pulsar appears to belong to a class of pulsars exhibiting fairly frequent 
occurrences of low amplitude glitches.  We calculate the glitch activity parameter for 
PSR J1833$-$1034 to be $1.53\times10^{-15}$ $s^{-2}$, which puts it in a special class of young 
pulsars like the Crab, and offset from the normal trend of glitch activity versus characteristic 
age (or spin frequency derivative) that a majority of the glitching pulsars follow.  This could be 
related to the thermal history of young neutron stars.

The final timing solution obtained after modeling of the glitches provides reliable estimates of 
the second derivative of the spin-down model for PSR J1833$-$1034.  The resulting braking index 
of 1.8569(6) is much less than the canonical value of 3, as also found for other young pulsars, 
supports the claim that pure dipole braking does not provide the full picture for pulsar spin-down.

With aid of the high time resolution and coherent dedispersion capabilities of the new 
GMRT Software Backend \citep{Roy10}, we aim to search of giant pulse (GP) emission from this young pulsar. 
Though it is thought that most of the GP emitters are neutron stars with strong magnetic 
field at the light cylinder ($B_{LC}$ = $10^{4}$ to $10^{5}$ G) \citep{Romani01}, the detection of
GPs in pulsars like J1752$+$2359 \citep{Ershov06}, B1112$+$50 \citep{Ershov03} and B0031$-$07 
\citep{Kuzmin04b,Kuzmin04a} reveals that GPs are also produced in pulsars with relatively 
low magnetic fields at the light cylinder. So even though the $B_{LC}$ of J1833-1034 is factor of 
7 lower than the Crab, it can be worth searching for GPs using the coherent dedispersed output taken 
with the GSB.  

\section{Acknowledgments}
The current work is based on 5.5 years of regular timing observations at the GMRT. We would like to thank all 
the staff members of the GMRT who are associated in maintaining and running of the telescope to make it available 
for the observations. We acknowledge the support of all the telescope 
operators who helped during these long extensive pulsar timing observations. The GMRT is run by the National 
Centre for Radio Astrophysics of the Tata Institute of Fundamental Research.  We would like to thank 
Prof. Dipankar Bhattacharya for insightful discussions on the theory of glitches.  We acknowledge the referee of this 
paper for his useful and constructive comments that helped to improve the quality of this paper significantly. Wojciech 
Lewandowski also acknowledges the support of the Polish Grant N N203 391934.

\label{lastpage}

\end{document}